# Composition of UHE Cosmic Ray Primaries

**J. Poirier, J. Carpenter, J. Gress, T. F. Lin, and A. Roesch**
*Physics Department, 225 NSH, University of Notre Dame, Notre Dame, IN 46556, USA*

## Abstract

Project GRAND presents results on the atomic composition of primary cosmic rays. This is accomplished by determining the average height of primary particles that cause extensive air showers detected by Project GRAND. Particles with a larger cross sectional area, such as iron nuclei, are likely to start an extensive air shower higher in the atmosphere whereas protons, with a smaller cross section, would pass through more air before interacting and thus start showers at lower heights. Such heights can be determined by extrapolating identified muon tracks backward (upward) to determine their height of origin (Gress et al., 1997). Since muons are from the top, hadronic part of the shower, they are a good estimator for the beginning of the shower. The data for this study were taken during the previous year with 20 million shower events.

Project GRAND is an extensive air shower array, located about 150 miles east of Chicago, USA (Poirier et al., 1999). It detects charged secondary particles from cosmic ray primaries by means of sixty-four tracking stations of four proportional wire chambers (PWCs), each containing independent x and y planes. If the top three PWCs in a hut detect particles within 0.1 microseconds, the data from all the wires in that hut at that time are temporarily stored in the hut. The presence of a track from that hut is reported to the main computer which is centrally located in the array. If at least four of the sixty-four huts report a track at the same time, the time of the coincidence and the huts which participated in the coincidence are recorded into temporary memory, and the status of the 41,000 wires from the 64 huts is downloaded to the main computer and stored on magnetic tape.

A 50 mm steel plate above the bottom chamber allows muon tracks to be identified (Kochocki et al., 1991). Pattern recognition can be used to identify muon candidates since muons pass through the steel plate with minimal deflection whereas electrons normally stop, shower, or deflect by large angles. A set of hits is considered to be a muon candidate if the hits in the second and third PWCs are consistent to within two cells with a line fit to the first and fourth hits in a given projection and if there are no hits other than those that comprise the track within two cells of the tracks in the top three PWCs or within five cells in the fourth chamber. This cut is wider in the fourth chamber to reject neighboring electron tracks that shower or scatter in the steel. Once a muon candidate is chosen, a line is fit to the hits in the top three detectors. A pseudo scattering angle is determined as the angular difference between the predicted hit position and the actual hit position of the candidate particle. Tracks are accepted as muons only if the pseudo scattering angle is less than 6° ensuring that a mere 2.7% of electron tracks are misidentified as muons as measured in a subsidiary experiment (Gress et al., 1991).

In the next step in the analysis, the software determines the position and size of the shower. The shower core location is determined by maximizing a log-likelihood function (LLF) of the number of particles detected in each station for a given shower. The most likely core location on an 8 x 8 grid is determined. A quadratic fit is performed with nearest grid neighbor's LLF to interpolate the location of the shower core in the grid. Then the shower core is required to lie inside the detector array by at least seven meters from any edge. Once the shower core is found to be within the array, the best fit function is integrated inside a 100m x 100m area to find the total number of charged particle hits within this area. A bounded area is used because an integration to infinity is slightly unstable and very dependent on the model used to extrapolate to large distances. The totals for this bounded area are almost model independent.

After the program completes determining the number of charged tracks they are put in size groups: <300, 300-1000, 1K-3K, 3K-10K, 10K-30K, 30K-100K, >100K. The program then sums all the muon angles of the tracks of each shower and averages them in x and y projection to find shower core direction. Almost verticle showers were selected by demanding the shower's direction to be within 30 degrees of zenith in the xz and yz planes. The program goes through all identified muon tracks of the shower and calculates the separation of all pairs and their difference angle. The calculation is conducted separately in the xz and yz planes. This difference angle, $\Delta\theta$, is averaged for each x or y interval of separation. In order to discard background muon tracks, the absolute value of the difference angle is required to be less than 6.7°.

From the track produced by each charged particle, its angle from zenith is calculated. These angles are measured with an average angular resolution of 0.26 degrees for each track. Knowing $\Delta\theta$ and the x or y separation between them, d, the height of the primary particle can be found from geometry and a small angle approximation

$$h = d / \Delta\theta \qquad (1)$$

where d and $\Delta\theta$ are projections either in the xz plane or yz plane. The height, h, is directly related to the composition of the primary cosmic ray. Particles with a larger cross section, such as iron nuclei, are more likely to interact sooner (at a greater height) than particles with smaller cross sections, such as protons.

If the composition of the primary cosmic rays is constant over energies which comprise the various shower size ranges, the height will decrease at a constant rate as the shower size increases. The number of charged particles from a shower that hit the detectors of Project GRAND are used to calculate the size of the shower within 100m x 100m. Figure 1 shows the mean angle for each of six shower size ranges versus the hut separation. The mean height is then determined from the mean angles and the distance between huts for each separation. For each shower size range, Figure 2 displays the mean muon height versus the hut separation. For a fixed height, equation (1) shows that the separation and mean angle must increase proportionally. However, Monte Carlo data indicate that the height is not constant but increases slightly with greater separation (Poirier et al., 1995). Thus, the mean angle should increase less than linearly with separation. Height, h, as a function of hut separation, x, for each shower size range thus is fit to a linear equation

$$h(x) = a + bx$$

in order to allow a change in height with increasing separation. The variable, x, represents either the x or y separation in meters. The origin of x is at a separation of 3.5 huts or 49m. In this equation, the parameters of 'a' and 'b' and their errors are varied to obtain a minimum $\chi^2$ value for each range of shower size. Table 1 lists the values of 'a' and 'b' and the minimum $\chi^2$ for each range of shower size.

**Table 1:** Fitted parameters to the function h(x) = a + bx, where a is the height centered in the array in meters and b is the slope for specific ranges of shower size by minimizing the $\chi^2$.

| Size of shower | a (m) | b | $\chi^2$ |
| --- | --- | --- | --- |
| 300 - 1000 | 5711 +/- 91 | 6 +/- 4 | 1.7 |
| 1K - 3K | 5207 +/- 42 | 13 +/- 2 | 0.5 |
| 3K - 10K | 4588 +/- 34 | 16 +/- 2 | 3.2 |
| 10K - 30K | 4467 +/- 53 | 5 +/- 2 | 8.6 |
| 30K - 100K | 4259 +/- 98 | -3 +/- 3 | 2.7 |
| >100K | 3859 +/- 339 | -1 +/- 10 | 2.2 |

Figure 3 plots 'a' as a function of the $\log_{10}$ of the shower size range. For these first three size groups there is a monotonic decrease in height similar to what is obtained in MC predictions for increasing energies as the shower penetrates deeper in the atmosphere (Poirier et al., 1995). There is a pronounced change at a shower size of 10K - 30K (charged particles inside 100m x 100m).

The point of increase in the spectral index as a function of the number of charged particles detected by Project GRAND can be seen in Figure 4. The nonstatistical fluctuations at 4.20 and 4.25 on the x-axis are most likely due to the coarse intervals of a table which correlates the number of wires hit to the number of tracks in the hut. The rapid change in the slope above ~100,000 charged particles (inside 100m x 100m area) of this graph is called the 'knee' -- the place where the rate of fall-off in cosmic ray flux increases at a certain energy [approximately proportional to the number of charged tracks; the actual energy value associated with a shower size (number) depends on the assumed composition]. It has been conjectured that the cause of the knee could perhaps be due to a change in composition near the energy of the knee. In Figure 3, evidence is seen for change (10K - 30K) in composition toward larger atomic numbers (heavier nuclei) which does indeed seem to be correlated with the knee observed at a nearby value (~100K) in Figure 4. Both features can be qualitatively understood since the lighter atoms (like protons) have a smaller charge, they have a larger gyroradius in the galaxy's magnetic field and hence have an increasingly better chance to escape as their momentum (and gyroradius) increases to values larger than the dimensions of coherent magnetic field regions in the galaxy. If the lighter atoms begin to escape, the composition would change to heavier, on the average, and, as well, cause an increase in the spectral slope.

Project GRAND is presently being funded by grants from the University of Notre Dame and private donations.

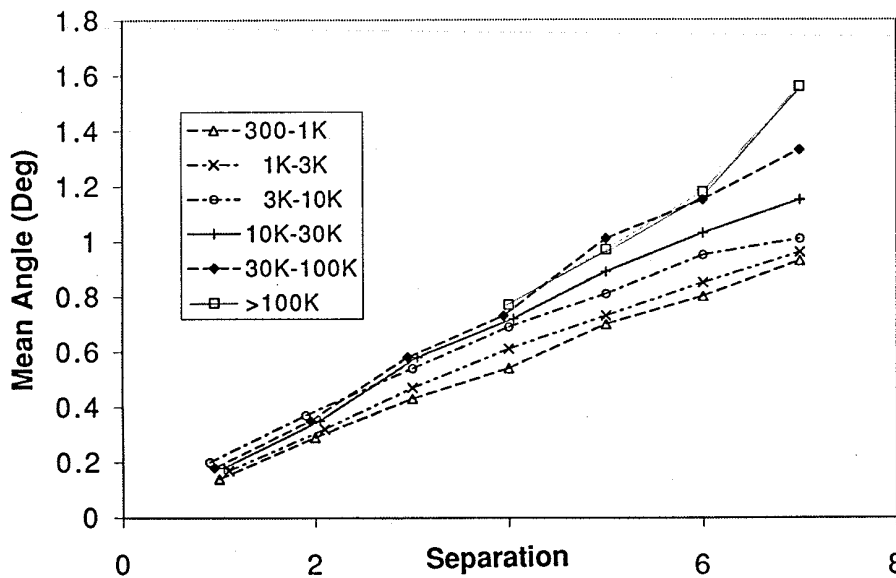

**Figure 1:** Mean angular difference between muon tracks vs hut separation (14 meters between huts) for each of six shower size ranges (number of charged tracks within 100m x 100m).

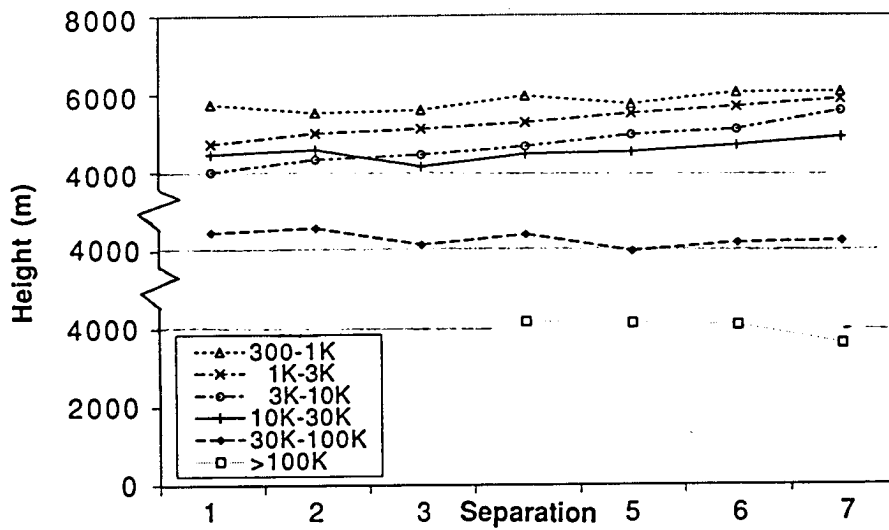

Figure 2: Mean height of muon origin in meters vs hut separation (14 meters between huts) for each of six shower size ranges (number of charged tracks within 100m x 100m). The two highest size ranges were separated to eliminate overlap. Lines at 4000m help orient the displaced values.

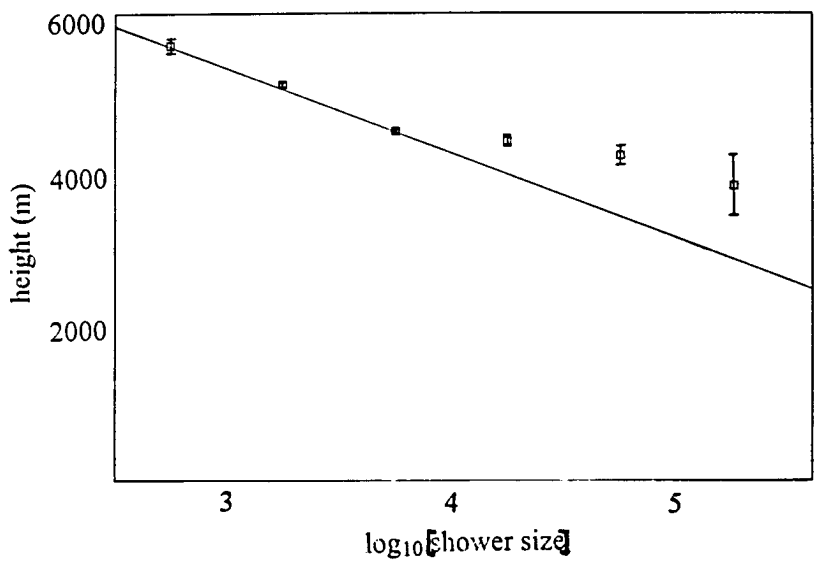

Figure 3: Fitted height in meters vs $\log_{10}$[shower size] (number of charged tracks within 100m x 100m).

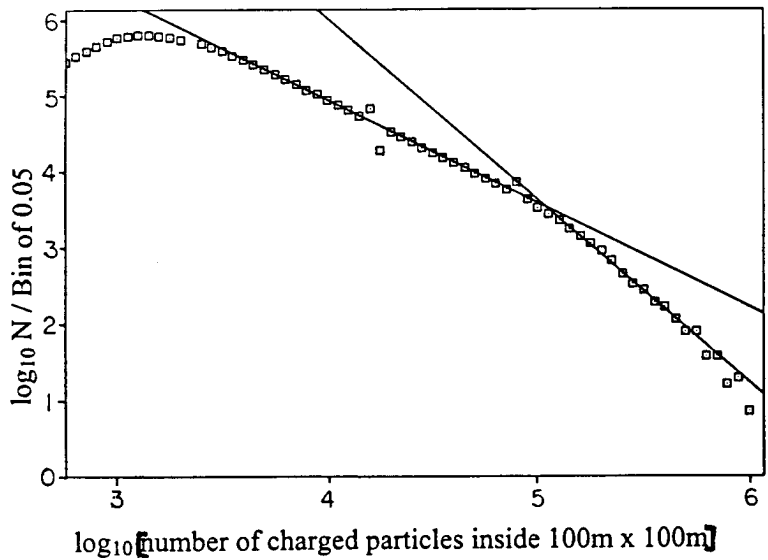

Figure 4: Histogram of $\log_{10}$ of the number of showers vs $\log_{10}$ their size. "Size" is the number of charged tracks inside 100m x 100m. A "knee" is observed at ~100,000 charged tracks inside the bounded area.